\documentclass[12pt,twocolumn,preprint]{aastex6}
\UseRawInputEncoding 
\usepackage{CJK}
\usepackage{color}

\usepackage{natbib}
\usepackage{hyperref}
\usepackage{cleveref}
\usepackage{url}

\begin{document}
\title{Impact of NICER's Radius Measurement of PSR J0740+6620 on Nuclear Symmetry Energy at Suprasaturation Densities}
\author{Nai-Bo Zhang\altaffilmark{1} and Bao-An Li\altaffilmark{2}$^{*}$}
\altaffiltext{1}{School of Physics, Southeast University, Nanjing 211189, China}
\altaffiltext{2}{Department of Physics and Astronomy, Texas A$\&$M University-Commerce, Commerce, TX 75429, USA\\
\noindent{$^{*}$Corresponding author: Bao-An.Li@Tamuc.edu}}

\begin{abstract}
By directly inverting several neutron star observables in the three-dimensional parameter space for the Equation of State of super-dense neutron-rich nuclear matter, we show that the lower radius limit for PSR J0740+6620 of mass $2.08\pm 0.07~M_{\odot}$ from Neutron Star Interior Composition Explorer (NICER)'s very recent observation sets a much tighter lower boundary than previously known for nuclear symmetry energy in the density range of $(1.0\sim 3.0)$ times the saturation density $\rho_0$ of nuclear matter. The super-soft symmetry energy leading to the formation of proton polarons in this density region of neutron stars is clearly disfavoured by the first radius measurement  for the most massive neutron star observed reliably so far.
\end{abstract}
\maketitle

\section{Introduction}
The PSR J0740+6620 having an updated mass of $2.08\pm 0.07~M_{\odot}$ \citep{Fonseca21} remains the most massive neutron star (NS) discovered so far \citep{Mmax}. Its radius of $13.7^{+2.6}_{-1.5}$ km (68\%) \citep{Miller21} or $12.39_{-0.98}^{+1.30}$\,km \citep{Riley21} was very recently inferred in two independent analyses of the X-ray data taken by the \textit{Neutron Star Interior Composition Explorer} (NICER) and the X-ray Multi-Mirror
(XMM-Newton) observatory. Combined with NICER's earlier simultaneous mass and radius measurement of PSR J0030+0451 \citep{Miller2019,Riley2019}, the first radius measurement of the most massive NS has the strong potential to reveal interesting new physics about the Equation of State (EOS) of super-dense neutron-rich nuclear matter. Besides earlier predictions about what uniquely new physics can be learned from the radii of massive NSs compared to canonical ones, see, e.g., \citet{Xie20,Han20,Dris21,Som21}, several new analyses aiming at extracting new information about the EOS of super-dense matter from NS observations including the latest NICER observations have already been carried out \citep{Biswas21,Li21,Raa21,Pang21,Essick21,Huth21}. Most of these studies are based on the Bayesian statistical inference. Since the approximately 10-15\% statistical uncertainty on top of the roughly 10\% systematic uncertainty (as indicated by the difference of the results from the two analyses) of NICER's radius measurements are still quite large, improvements to the existing constraints of nuclear EOS brought by the latest NICER measurement were found to be generally small within the Bayesian statistical analyses. 

After verifying the small effects of NICER's new radius measurements on the EOS in a Bayesian analysis \citep{XieLi21}, here we perform a direct inversion of the lower radius boundaries obtained by both \citet{Miller21} and \citet{Riley21} in a three-dimensional (3D) high-density EOS parameter space. Without the statistical averaging over the whole uncertainty range of observables normally done in the Bayesian analyses, direct inversion of specific values of NS observables can reveal interesting features that are generally smeared out in the Bayesian statistical analyses. We skip the inversion of the upper radius boundaries from the two NICER analyses as they do not provide any tighter constraint on the EOS compared to what is available from analyzing the upper limit of tidal deformation of GW170817. We show that the 68\% lower mass-radius boundaries from the two independent analyses by \citet{Miller21} and \citet{Riley21} provide a much tighter lower boundary than previously known for nuclear symmetry energy in the density range of $(1.0\sim 3.0)\rho_0$, disfavoring the super-soft symmetry energy necessary for the formation of proton polarons in this density region in NSs and the associated phenomena predicted in the literature.

The rest of the paper is organized as follows. In the next section, we recall the main challenges and ramifications of determining nuclear symmetry energy at suprasaturation densities.  We then
summarize very briefly the direct inversion approach of analyzing NS observables in the 3D high-density EOS parameter space. We will demonstrate in Section \ref{Results} the scientific power of NICER's radius measurement of PSR J0740+6620 as the most massive NS with a reliable mass in tightening the lower limit of high-density nuclear symmetry energy. Effects of the remaining uncertainty of the slope parameter $L$ of nuclear symmetry energy will be discussed in Section \ref{Sec4}. Finally, we summarize our main findings.

\section{The challenges and ramifications of determining nuclear symmetry energy at suprasaturation densities}
The average energy per nucleon $E(\rho ,\delta )$
in neutron-rich matter of nucleon density $\rho=\rho_n+\rho_p$ and isospin asymmetry $\delta\equiv (\rho_n-\rho_p)/\rho$ can be written as \citep{Bom91}
\begin{equation}\label{eos}
E(\rho ,\delta )=E_0(\rho)+E_{\rm{sym}}(\rho )\cdot \delta ^{2} +\mathcal{O}(\delta^4)
\end{equation}
 where $E_{0}(\rho)$ is the nucleon energy in symmetric nuclear matter (SNM) while $E_{\rm{sym}}(\rho)$ is the nuclear symmetry energy. The latter measures the energy cost to make nuclear matter more neutron-rich. It is essentially the difference in energy per nucleon in pure neutron matter (PNM) and SNM.
 The $E_{\rm{sym}}(\rho)$ is very poorly known especially at suprasaturation densities mainly because of our poor knowledge about the spin-isospin dependence of three-body nuclear forces, the isospin dependence of the tensor force, and the related short-range nucleon-nucleon correlations in dense neutron-rich matter \citep{Li18,BALI19}. It has been well known that theoretical predictions of high-density $E_{\rm{sym}}(\rho)$ diverge broadly at suprasaturation densities, ranging from large negative to positive values at densities above about $(2\sim 3)\rho_0$ \citep{Li98,Steiner05,Ditoro,LCK08,Tesym,Bal16}. While there have been continuously a lot of new efforts and indeed some impressive progresses in recent years in predicting nuclear symmetry energy at densities below $2\rho_0$ using microscopic nuclear many-body theories, the situation at higher densities remain essentially the same, see, e.g., the summary in Figure 2 in the latest review \citep{Bur21} or Figure 8 in \citep{Li21}. To our best knowledge, there is no fundamental physics principle forbidding the $E_{\rm{sym}}(\rho)$ to become zero or even negative at suprasaturation densities. In fact, based on variational nuclear many-body theory calculations \citep{Panda,WFF12}, it was pointed out already that when the repulsive short-range tensor force due to the $\rho$ meson exchange in the isosinglet nucleon-nucleon interaction channel in SNM becomes dominating at high densities, the energy in SNM increases faster than that in PNM, leading to a decreasing $E_{\rm{sym}}(\rho)$ with increasing density. The $E_{\rm{sym}}(\rho)$ may then become zero or even negative above certain critical densities as the $E_{\rm{sym}}(\rho)$ is approximately the energy difference between PNM (where the tensor force in the isotriplet nucleon-nucleon interaction channel is known to be negligible) and SNM. Indeed, it was demonstrated quantitatively by \citet{Xu1,Xu2} that whether the $E_{\rm{sym}}(\rho)$ increases or decreases and when it may become negative depend strongly on properties of the three-body force, the strength and short-range cut-off of the tensor force as well as the related isospin dependence of short-range nucleon-nucleon correlations.

The high-density behavior of $E_{\rm{sym}}(\rho)$ has significant ramifications in both nuclear physics and astrophysics \citep{Kut93,Kut94,Szm06,Ditoro,LCK08,Ditoro2,Bal16,YZhou19}. For example, variations of nuclear symmetry energy can lead to large changes in the binding energy and spacetime curvature near the surface of NSs \citep{Newton09,He15}. To further test Einstein's General Relativity in the strong-field gravity regime against modified gravity theories or the possible existence of a weakly interacting light boson mediating a new force using massive NSs thus requires a reliable knowledge of the high-density $E_{\rm{sym}}(\rho)$ \citep{Kri09,Wen09,Wlin14,Jiang15}. Moreover, the density profile of proton fraction in NSs at $\beta$-equilibrium is uniquely determined by the density dependence of nuclear symmetry energy and has many consequences. For instance, it determines the critical density above which the direct URCA process responsible for the fast cooling of protoneutron stars can happen. In particular, a super-soft $E_{\rm{sym}}(\rho)$, i.e., a decreasing/negative symmetry energy at high densities may lead to some very interesting new phenomena in the core of NSs. As the symmetry energy decreases with increasing density, the proton fraction decreases correspondingly, leading to the formation of proton polarons (isolated and localized single or small clusters of protons surrounded by high-density neutrons) \citep{Kut93,Kut94} in the core of NSs. It can then strongly affect transport coefficients of NS matter and can
produce spontaneous magnetization in NSs \citep{Szm06}. When the $E_{\rm{sym}}(\rho)$ becomes zero at very high densities in the core of NSs, matter there are completely PNM. Since we do not have a physical mechanism to handle negative symmetry energy in NSs in the present work, we use a cut-off at zero symmetry energy at high densities in cases where the  $E_{\rm{sym}}(\rho)$ continuously decreases towards negative values. Nevertheless, we notice that a negative symmetry energy and its effects on NSs have been studies by others, see, e.g., \citet{Kut93,Kut94,Szm06}. In particular, it has been shown that a negative $E_{\rm{sym}}(\rho)$ has dramatic effects on the possible kaon condensation in the cores of NSs \citep{Kubis1}. Theoretically, in nuclear matter (not necessarily in NSs), if the $E_{\rm{sym}}(\rho)$ becomes negative at super-high densities, the so-called isospin separation instability happens \citep{Kut94,Li02}. Namely, because of the $E_{\rm{sym}}(\rho)\cdot\delta^2$ term in the average nucleon energy of Equation (\ref{eos}) in isospin-asymmetric nuclear matter, a uniform symmetric nuclear matter ($\delta=0$) of energy $E_0(\rho)$ is unstable energetically against being separated into regions of bulk pure neutron matter ($\delta=+1$) and pure proton matter ($\delta=-1$) when $E_{\rm{sym}}(\rho)$ is negative because then the $E_{\rm{sym}}(\rho) (\delta=-1)^2 + E_{\rm{sym}}(\rho) (\delta=+1)^2 $ will be negative instead of zero, making the total energy of the system less than the $E_0(\rho)$ that a uniform SNM normally has.

To pin down the high-density behavior of $E_{\rm{sym}}(\rho)$ has been a major scientific thrust of high-energy rare isotope beam facilities \citep{LRP2015,NuPECC}, see, e.g., \citet{SEP,GSI} for examples of current and planned nuclear reaction experiments probing the high-density $E_{\rm{sym}}(\rho)$. In fact, there were some circumstantial evidences for a super-soft $E_{\rm{sym}}(\rho)$ at $\rho\geq 1.5\rho_0$ from analyzing data of an earlier heavy-ion reaction experiment \citep{XiaoPRL}. However, the conclusion remains controversial \citep{Feng12,Feng12b,Xie14,MSU} largely due to the strong model dependence and the relatively small isospin asymmetry reached in heavy-ion reactions \citep{Li17}. Thus, a tight lower boundary for the high-density $E_{\rm{sym}}(\rho)$ from measuring the radii of massive NSs are scientifically invaluable for both nuclear physics and astrophysics.

While the radii and tidal deformabilities of canonical NSs provide some useful constraints on the $E_{\rm{sym}}(\rho)$ around $(1\sim 2)\rho_0$, they do not constrain significantly the $E_{\rm{sym}}(\rho)$ at densities above about $2\rho_0$ as demonstrated clearly already in \citet{Zhang19,Xie19}. This is mainly because the radii and tidal deformabilities of canonical NSs are determined by the nuclear pressure around $(1\sim 2)\rho_0$ in these relatively light NSs \citep{Lattimer01}. Moreover, the pressure in this density region is strongly controlled by the density dependence of nuclear symmetry energy, see, e.g., Figure 5 in \citet{Li-mex} for a numerical demonstration. The radius measurement of canonical NSs thus constrains mainly the symmetry energy in this density region. At higher densities, the pressure is dominated by the SNM pressure. As pointed out already by \citet{Lattimer01}, ``there is a quantitative relation between the radius and the pressure that does not depend upon the EOS at the highest densities, which determines the overall softness or stiffness (and hence, the maximum mass)". Nowadays, when analyzing NS observables to extract information about nuclear EOS, one always first applies the prerequisite that all EOSs have to support the currently observed maximum mass of NSs. Under this condition, the radius data of canonical NSs thus has little effect on the SNM EOS in the whole density region but mainly constrain the $E_{\rm{sym}}(\rho)$ below about $2\rho_0$. At densities above $2\rho_0$, the remaining gap between the extracted upper and lower limits of $E_{\rm{sym}}(\rho)$  remains very large, see, e.g., Figure 8 in the recent review in \citet{Li21}, because the constrains on the $E_{\rm{sym}}(\rho)$ obtained from the radii of canonical NSs can only rule out some extreme cases predicted by some theories. In fact, it was shown clearly before the GW170817 was discovered that to probe the high-density behavior of $E_{\rm{sym}}(\rho)$ one has to use the radii and/or tidal deformabilities of massive NSs while those of the canonical ones only probe the $E_{\rm{sym}}(\rho)$ around $\rho_0$, see, e.g., \citet{FF1,FF2}. NICER's very recent radius measurement of PSR J0740+6620 made this hope a reality for the first time.

\begin{figure*}
  \centering
   \resizebox{1.0\textwidth}{!}{
  \includegraphics{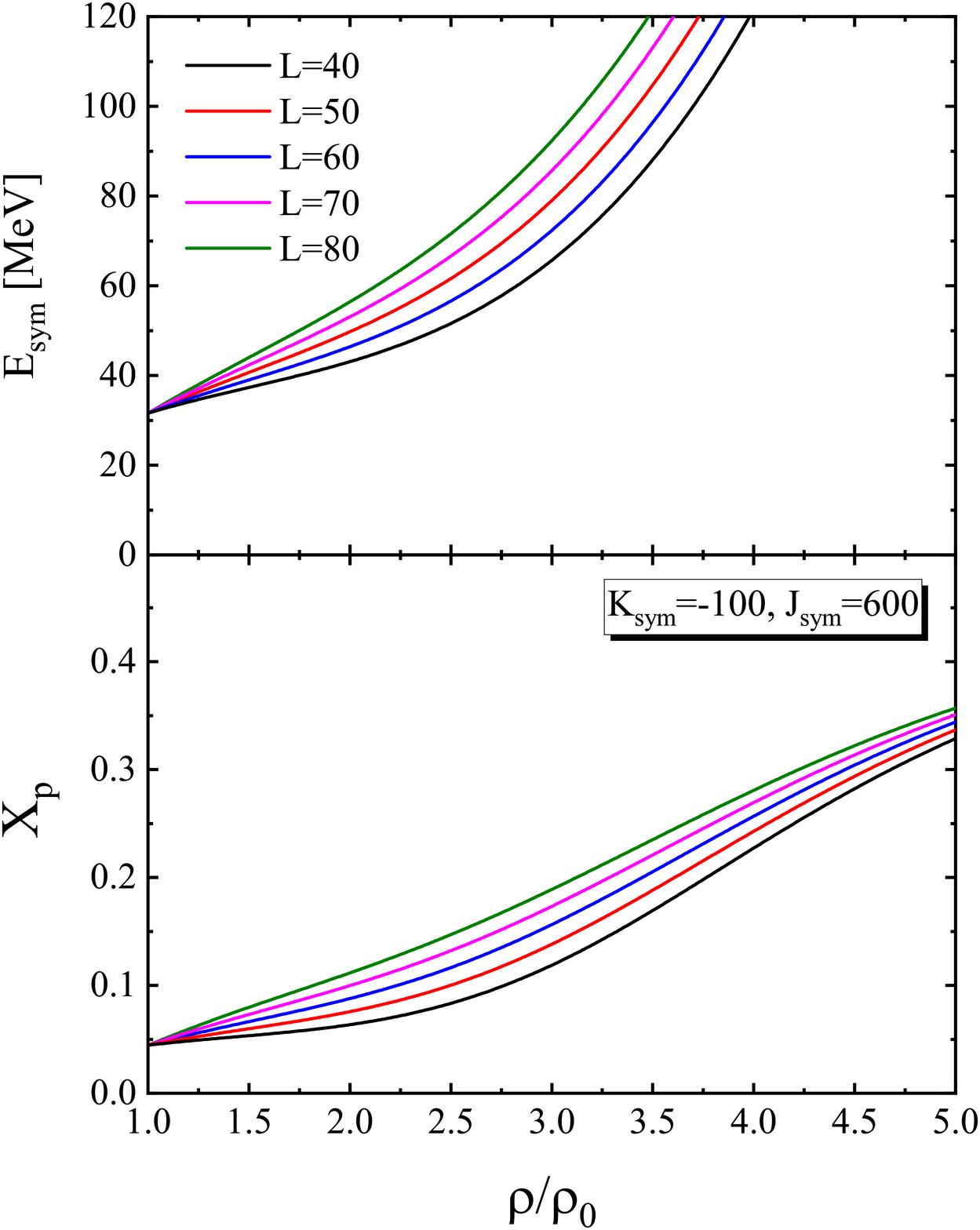}
  \includegraphics{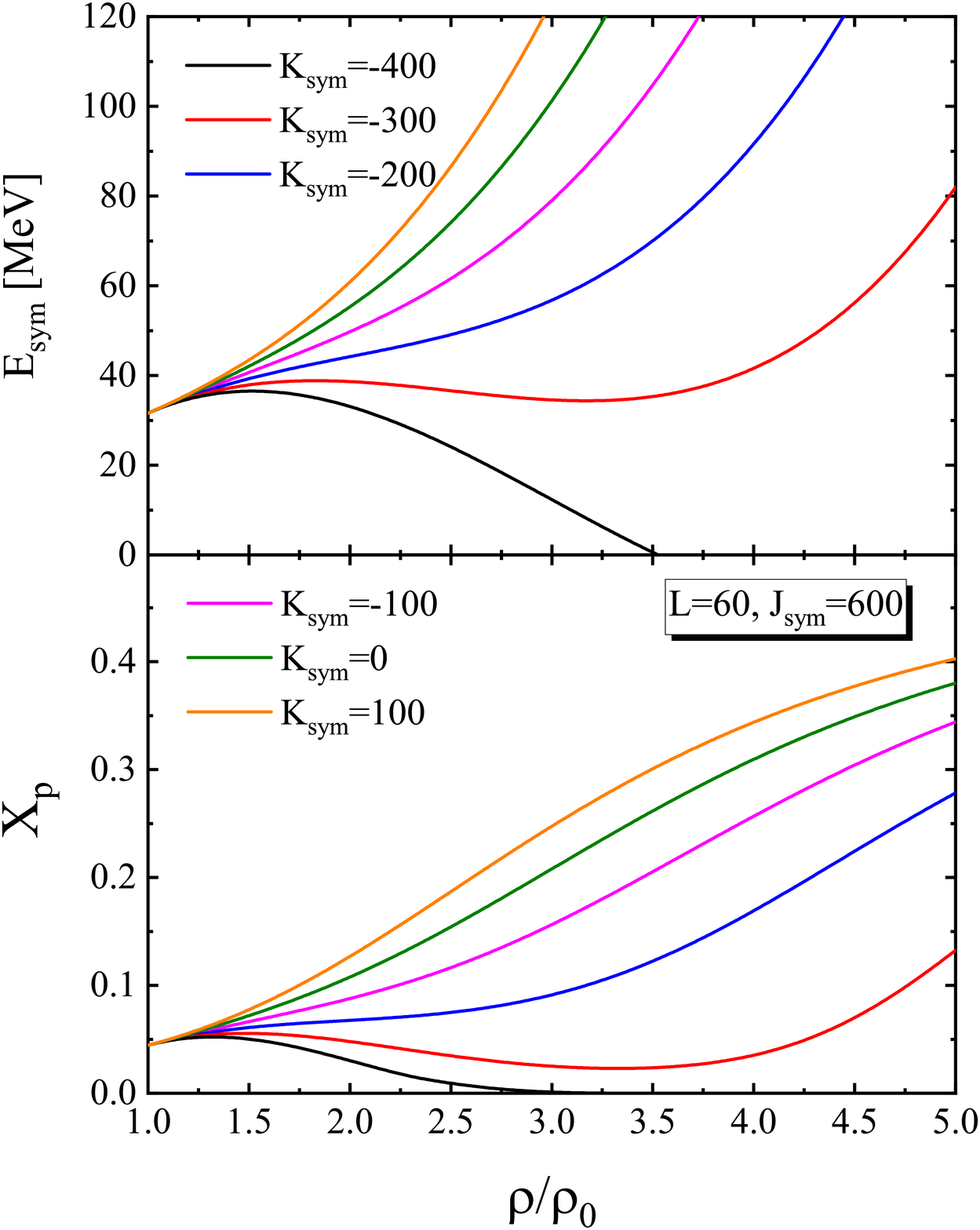}
  \includegraphics{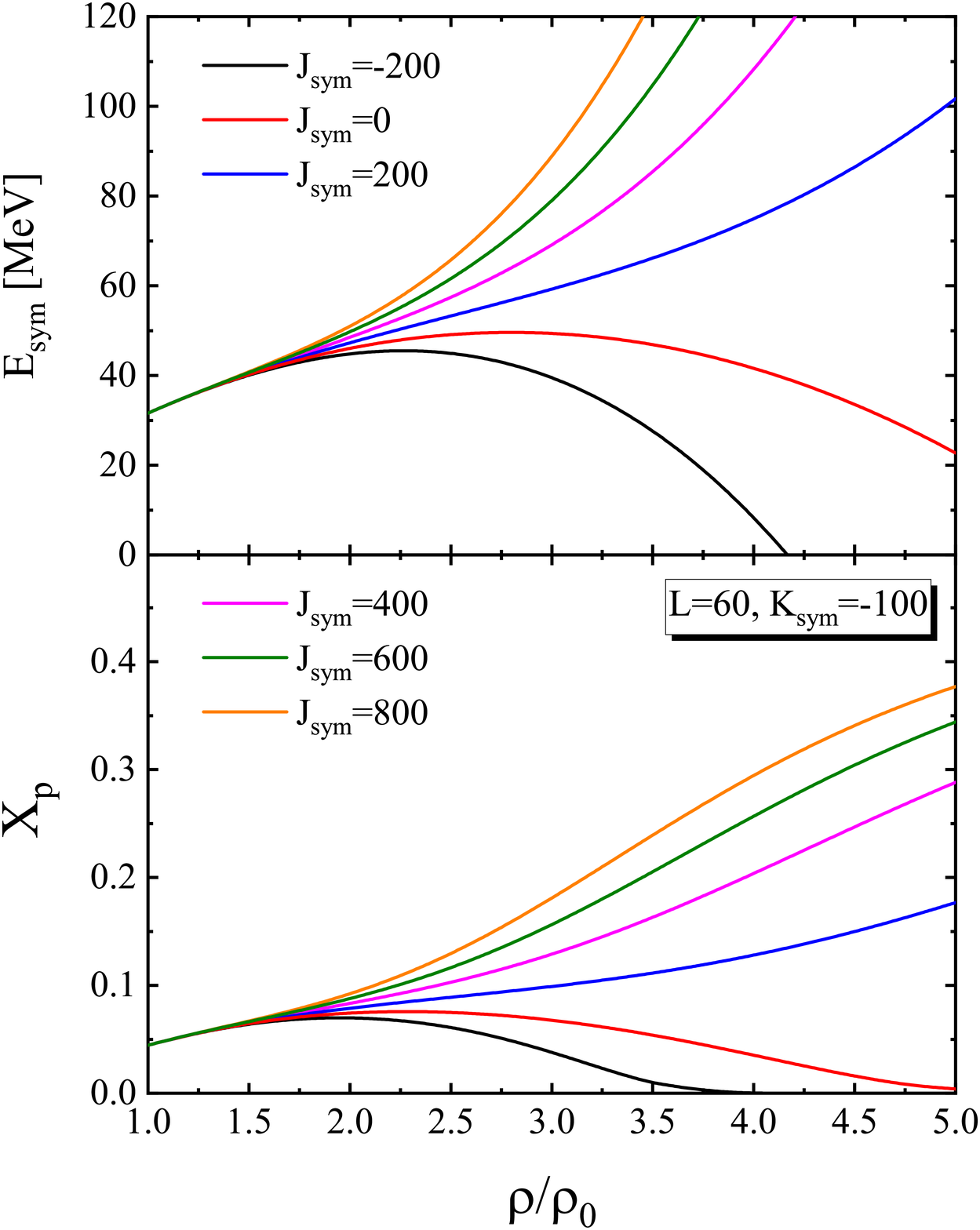}
   }
  \caption{The symmetry energy (upper panels) and proton fraction (lower panels) as functions of density from varying $L$ (left panel), $K_{\rm sym}$ (middle panel), and $J_{\rm sym}$ (right panel), respectively.}\label{Esymplot}
\end{figure*}
\section{Direct inversion of neutron star observables in high-density EOS parameter space}\label{EOSmodel}
For completeness and ease of our discussions, here we summarize briefly the main features of the NS EOS-metamodel we use in solving the NS inverse-structure problem. More details can be found in our previous publications
\citep{Zhang18,Zhang19,Zhang19a,Zhang19b,Zhang2020,Xie19,Xie20}.
We assume the cores of NSs are made of totally charge neutral neutrons, protons, electrons, and muons (the $npe\mu$ model) at $\beta$-equilibrium. Unlike the widely used composition-degenerate spectral functions and/or piecewise polytropes that directly parameterize the pressure as a function of energy or baryon density, to probe the high-density symmetry energy we have to keep the isospin dependence of the EOS and retain explicitly the composition information at all densities. For this reason, we metamodel NS EOS by starting at the single nucleon energy level with explicitly isospin dependence. Specifically, we parameterize the $E_{0}(\rho)$ and $E_{\rm{sym}}(\rho)$ according to
\begin{eqnarray}\label{E0para}
  E_{0}(\rho)&=&E_0(\rho_0)+\frac{K_0}{2}(\frac{\rho-\rho_0}{3\rho_0})^2+\frac{J_0}{6}(\frac{\rho-\rho_0}{3\rho_0})^3,\\
  E_{\rm{sym}}(\rho)&=&E_{\rm{sym}}(\rho_0)+L(\frac{\rho-\rho_0}{3\rho_0})\nonumber\\
  &+&\frac{K_{\rm{sym}}}{2}(\frac{\rho-\rho_0}{3\rho_0})^2
  +\frac{J_{\rm{sym}}}{6}(\frac{\rho-\rho_0}{3\rho_0})^3\label{Esympara}
\end{eqnarray}
where $E_0(\rho_0)=-15.9 \pm 0.4$ MeV is the binding energy and $K_0\approx 230 \pm 20$ MeV \citep{Shlomo06,Piekarewicz10,Garg18} is the incompressibility at the saturation density $\rho_0$ of SNM, while $E_{\rm sym}(\rho_0)=31.7\pm 3.2$ MeV is the magnitude and $L\approx 58.7\pm 28.1 $ MeV is the slope of symmetry energy at $\rho_0$ \citep{Li13,Oertel17} based on earlier surveys of over 50 analyses of both terrestrial experiments and astrophysical observations, respectively. A very recent survey of 24 new analyses of NS observations since GW179817 indicates that $L\approx 57.7\pm 19$ MeV and $K_{\rm{sym}}\approx -107\pm 88$ MeV at 68\% confidence level \citep{Li21}. In this study, we use $L=58.7$ as its most probable value and vary it within $\pm20$ MeV. We keep the $E_0(\rho_0), K_0$, and $E_{\rm sym}(\rho_0)$ at their most probable values given above as they have been relatively well determined and their variations within their remaining uncertain ranges have been shown to have little effects on the masses, radii, and tidal deformabilities of NSs \cite{FF1,FF2,Xie19,Xie20}.
As a reference, we notice that very recent calculations based on a set of commonly used Hamiltonians including two- and three-nucleon forces derived from chiral effective field theory predicted that $L=47\pm3$ MeV, 
 $K_{\rm{sym}}= -146\pm 43$ MeV, $J_{\rm{sym}}=90\pm 334$ MeV \citep{Soma}. While in the same calculations, with a prior $J_0=-300 \pm 400$ MeV the posterior ranges between  $J_0=-573 \pm 133$ MeV and  $J_0=-172 \pm 243$ MeV depending on the scaling schemes used.  The EOS parameter ranges we used are consistent with these calculations. 

The $K_{\rm{sym}}$, $J_{\rm{sym}}$, and $J_0$ are parameters characterizing the EOS of super-dense neutron-rich nuclear matter. We directly invert NS observables in the high-density EOS space allowed by the current uncertainties of $K_{\rm{sym}}$, $J_{\rm{sym}}$, and $J_0$ as summarized in \citet{Zhang17}. We note qualitatively that while the parameter $J_0$ (skewness of SNM) controls the stiffness of SNM EOS at high densities, the $K_{\rm{sym}}$ (curvature of symmmetry energy) dominates the behavior of $E_{\rm sym}(\rho)$ around $2\rho_0$ and the parameter $J_{\rm{sym}}$ (skewness of symmetry energy) controls the $E_{\rm sym}(\rho)$ at densities above $(2\sim 3)\rho_0$. The slope parameter $L$ dominates the behavior of $E_{\rm sym}(\rho)$ around $\rho_0$ but does not have much effects on its high-density behavior. It is also known from previous Bayesian analyses and direct inversions that the $K_{\rm{sym}}$ and $J_{\rm{sym}}$ affect most strongly on the radii of massive NSs while the $L$ parameter influences most strongly the radii of canonical NSs \citep{Zhang19a,Xie20}.

The roles of the parameters $L$, $K_{\rm{sym}}$, and $J_{\rm{sym}}$ mentioned above are illustrated in Figure \ref{Esymplot} where the $E_{\rm sym}(\rho)$ (upper panels) and the corresponding proton fraction (lower panels) in NSs at $\beta-$ equilibrium are shown as functions of density by varying the slope $L$ (left panel), curvature $K_{\rm sym}$ (middle panel), and skewness $J_{\rm sym}$ (right panel), respectively. They can sample essentially all currently known variations of density dependence of nuclear symmetry energy. We loosely refer the $E_{\rm sym}(\rho)$ that quickly decreases to zero above $(2-3)\rho_0$ as ``super-soft". This terminology has been used by the nuclear physics community, see, e.g., \citet{XiaoPRL}. The $E_{\rm sym}(\rho)$ that increases much faster than linear is loosely referred as ``super-stiff", based on the fact that the original Relativistic Mean Field (RMF) model predicts a linear increase with density due to the $\rho-$meson exchange and the RMF prediction is often referred as being ``stiff". As shown in Figure \ref{Esymplot}, combinations of large negative (positive) $K_{\rm{sym}}$ and/or $J_{\rm{sym}}$ values lead to super-soft (super-stiff) symmetry energies at high densities. For a super-soft $E_{\rm{sym}}(\rho)$, the proton fraction $X_p$ quickly decreases to zero, indicating the formation of PNM. Conversely, the $X_p$ quickly approaches 1/2 indicative of a SNM at super-high densities for the super-stiff $E_{\rm{sym}}(\rho)$.

It is worth emphasizing that the Equations (\ref{E0para}) and (\ref{Esympara}) for $E_{0}(\rho)$ and  $E_{\rm{sym}}(\rho)$  have dual meanings that can be confusing.  If the functions $E_{0}(\rho)$ and  $E_{\rm{sym}}(\rho)$ are known {\it apriori} as in all forward-modelings and in some inversions using energy density density functionals, the Equations (\ref{E0para}) and (\ref{Esympara}) are Taylor expansions up to the third order in $\chi\equiv(\rho-\rho_0)/3\rho_0$. It is well known that the Taylor expansion may not converge at high densities. In fact, it has been shown very recently in \citet{CaiLi21} that even if all the coefficients in the Equations (\ref{E0para}) and (\ref{Esympara}) are well determined by whatever experiments and/or theories, the Taylor expansions even up to $\chi^5$ can not reproduce accurately the known $E_{0}(\rho)$ and  $E_{\rm{sym}}(\rho)$ at densities around $(3-4)\rho_0$. To remedy this problem, an auxiliary function approach instead of the original Taylor expansion was proposed. However, it has not been implemented in any analysis of NS properties yet.

On the other hand, in our metamodeling of NS EOS in this work and our previous works using both Bayesian and direct inversion approaches,
the $E_{0}(\rho)$ and  $E_{\rm{sym}}(\rho)$ are {\bf NOT} known {\it apriori}. The Equations (\ref{E0para}) and (\ref{Esympara}) for $E_{0}(\rho)$ and $E_{\rm{sym}}(\rho)$ are simply parameterizations with their coefficients to be inferred directly from data. They are thus not Taylor expansions of any known functions. But they are purposely being parameterized in forms like Taylor expansions so that we can use existing knowledge about the coefficients of Taylor expansions of nuclear EOS as we mentioned above as guidances in setting the range and selecting the prior probability distribution functions (PDFs) of the parameters involved.
Mathematically, the parameterizations approach asymptotically to Taylor expansions at the limit of $\rho\rightarrow \rho_0$. Indeed, simply referring the parameters $L$, $K_{\rm sym}$, and $J_{\rm sym}$ as the slope, curvature, and skewness of symmetry energy is misleading. But neglecting this subtle difference has been a convenient and common practice in the literature.

Similar to the question of how many segments of polytropes one should use in parameterizing the EOS at high densities to achieve model independent results, to what order of $\chi$ one can practically use and how the results may be different are interesting questions. Some answers to these questions were given within the Bayesian approach in \citet{Xie19} and \citet{Xie21-JPG}. For example, the kurtosis parameter $Z_0$ is the coefficient of the $\chi^4$ term in Taylor expanding the SNM EOS $E_{0}(\rho)$. Its theoretical value is known roughly to be between -5330 MeV and +5038 MeV \citep{Jerome,Ant}. Using the empirical pressure in SNM from nuclear collective flow in relativistic heavy-ion collisions as data \citep{Pawel}, the most probable values and 68\% confidence boundaries of both the inferred $K_0$ and $J_0$ vary appreciable especially for the latter depending on whether or not the $Z_0$ term is included in parameterizing the $E_{0}(\rho)$ \citep{Xie21-JPG}. While the Bayesian inferred $Z_0$ value of $600\pm 200$ MeV is much more accurate than its prior value from the literature, its uncertainty remains very large and it affects mostly the SNM EOS above about $(3-4)\rho_0$. As another example, as demonstrated in Section 4 of \citet{Xie19}, the most probable values and confidence boundaries of $L$ and $K_{\rm sym}$ extracted from NS observables depend significantly on whether the $J_{\rm{sym}}$ and $J_0$ terms are considered in the EOS parameterizations. Indeed, in the analyses by different groups in the literature, often the $J_{\rm{sym}}$ is neglected as its range is very poorly known as we mentioned earlier. In fact, whether the $J_{\rm{sym}}$ is considered or not is partially responsible for the variations of $L$ and $K_{\rm sym}$ reported in the literature as pointed out already in \citet{Xie19}. As to the kurtosis parameter $Z_{\rm{sym}}$ of symmetry energy, it is currently known to be around $-1800\pm 800$ MeV \citep{Soma} and is not considered in our work. 

For visualization purposes, our direct inversion as we shall illustrate next is limited to 3D. Given the above information and for the purpose of extracting the $E_{\rm{sym}}(\rho)$ mostly in the region of $(2-4)\rho_0$, we perform the direct inversion in the $J_0-K_{\rm sym}-J_{\rm sym}$ 3D high-density EOS parameter space. For this purpose, the parameterizations of $E_{0}(\rho)$ and $E_{\rm{sym}}(\rho)$  are sufficient. It is not meaningful and impractical to include even higher order terms before the $J_{\rm sym}$ parameter is better determined. Using less terms is also not meaningful as the $J_{\rm sym}$ is most important for determining the $E_{\rm{sym}}(\rho)$ around $(2-4)\rho_0$. Thus, within the direct inversion approach we currently can not answer the question about how our results may change if more or less terms are used in parameterizing the EOS, and as we discussed earlier there is currently no theoretical consensus on the $E_{\rm{sym}}(\rho)$ above $2\rho_0$ for us to compare with. Therefore, all of our results should be understood within the context and cautions discussed above.

The pressure in the $npe\mu$ core of NSs is calculated from
\begin{equation}\label{pressure}
  P(\rho, \delta)=\rho^2\frac{d\epsilon(\rho,\delta)/\rho}{d\rho},
\end{equation}
where $\epsilon(\rho, \delta)=\epsilon_n(\rho, \delta)+\epsilon_l(\rho, \delta)$ denotes the total energy density with contributions from both nucleons ($\epsilon_n(\rho, \delta)$) and leptons ($\epsilon_l(\rho, \delta)$).
To see clearly where the symmetry energy comes in, we note in particular that the nucleon energy density $\epsilon(\rho,\delta)=\rho E(\rho,\delta)+\rho M$ with $M$ being the average mass of nucleons. Given the density dependence of nuclear symmetry energy $E_{\rm sym}(\rho)$, the density profile of isospin asymmetry $\delta(\rho)$ (or the corresponding proton fraction $x_p(\rho)$) is determined by the $\beta$-equilibrium and charge neutrality conditions
$
  \mu_n-\mu_p=\mu_e=\mu_\mu,~\rho_p=\rho_e+\rho_\mu
$
in terms of the chemical potential $\mu_i=\frac{\partial\epsilon(\rho,\delta)}{\partial\rho_i}$ and particle density $\rho_i$ of particle $i$.

We connect the core EOS described above with the NV EOS \citep{Negele73} for the inner crust and the BPS EoS \citep{Baym71b} for the outer crust. The crust-core transition density and pressure are determined consistently for each EOS generated for the core as discussed in detail in \citet{Zhang18}.

\begin{figure}
  \centering
   \resizebox{0.48\textwidth}{!}{
  \includegraphics[bb=0 0 550 570]{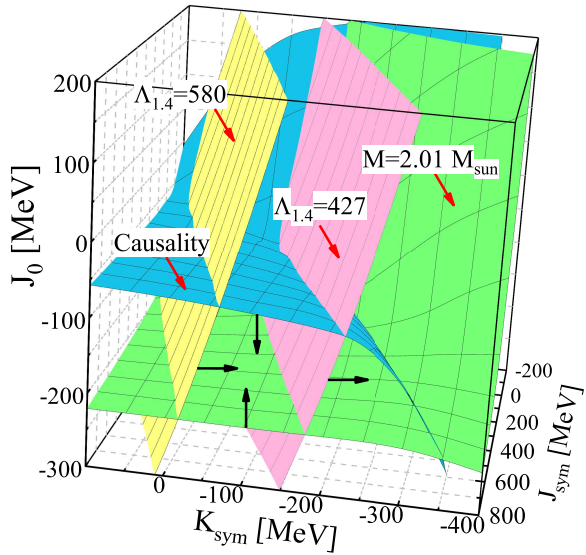}
  }
  \vspace{-1cm}
  \caption{Constant surfaces of NS observables and causality condition in the 3-dimensional $K_{\rm sym}-J_{\rm sym}-J_0$ EOS parameter space when $L$ is fixed at its currently known most probable value of $L=58.7$ MeV: the NS {\it minimum} maximum mass of $M=2.01$ M$_\odot$ (green surface), the dimensionless tidal deformability of canonical NS: $\Lambda_{1.4}=580$ at 90\% confidence level (yellow surface) and $\Lambda_{1.4}=427$ at 68.3\% confidence level (pink surface),  and the causality surface (blue) on which the sound speed equals the speed of light in centers of the most massive NSs. The red arrows point to the corresponding surfaces while the black arrows show directions that satisfy the corresponding constraints.}\label{NICER2L60LIGO}
\end{figure}
\begin{figure*}
  \centering
  \resizebox{1.2\textwidth}{!}{
     \includegraphics{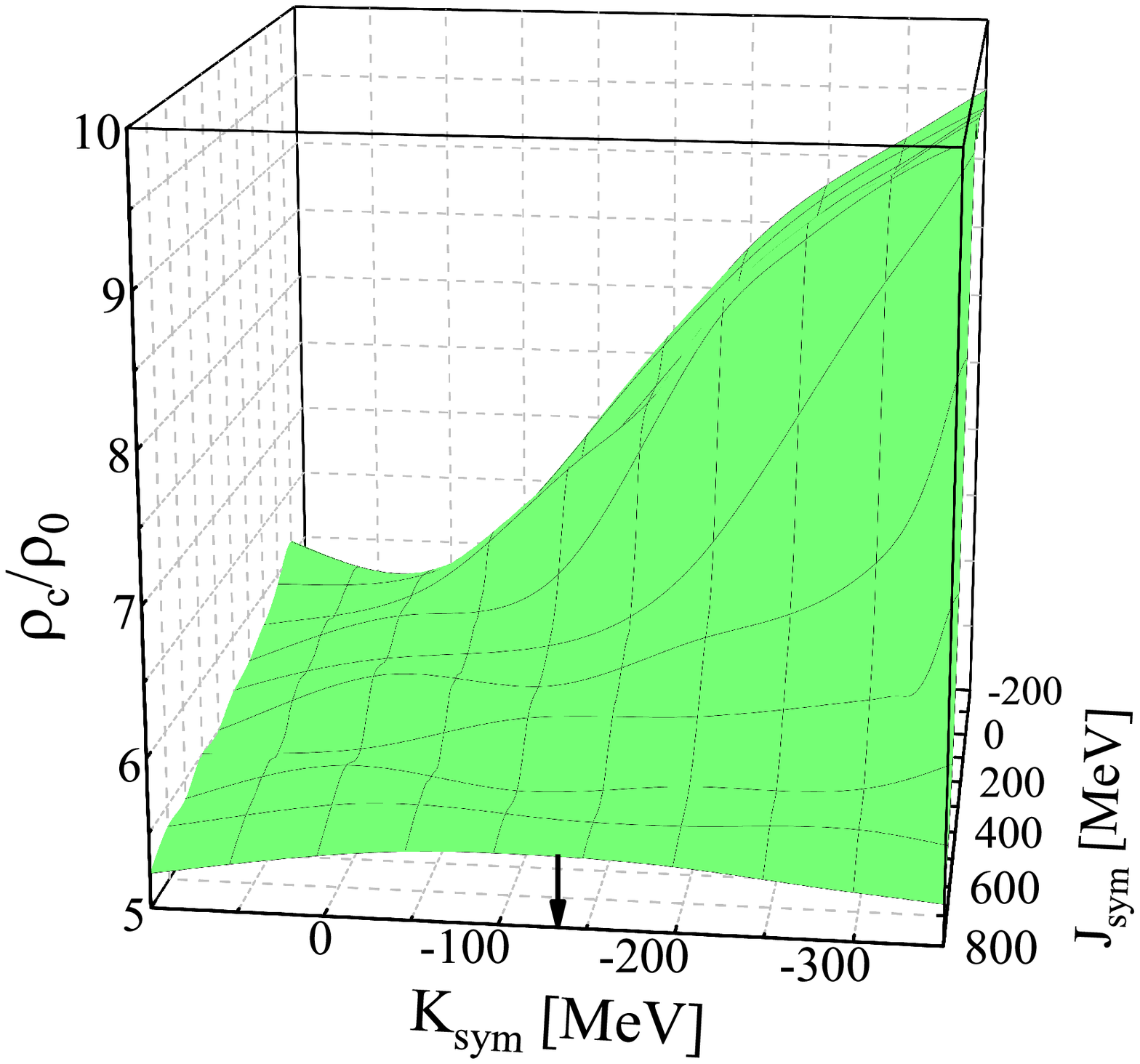}
     \hspace{-6cm}
  \includegraphics{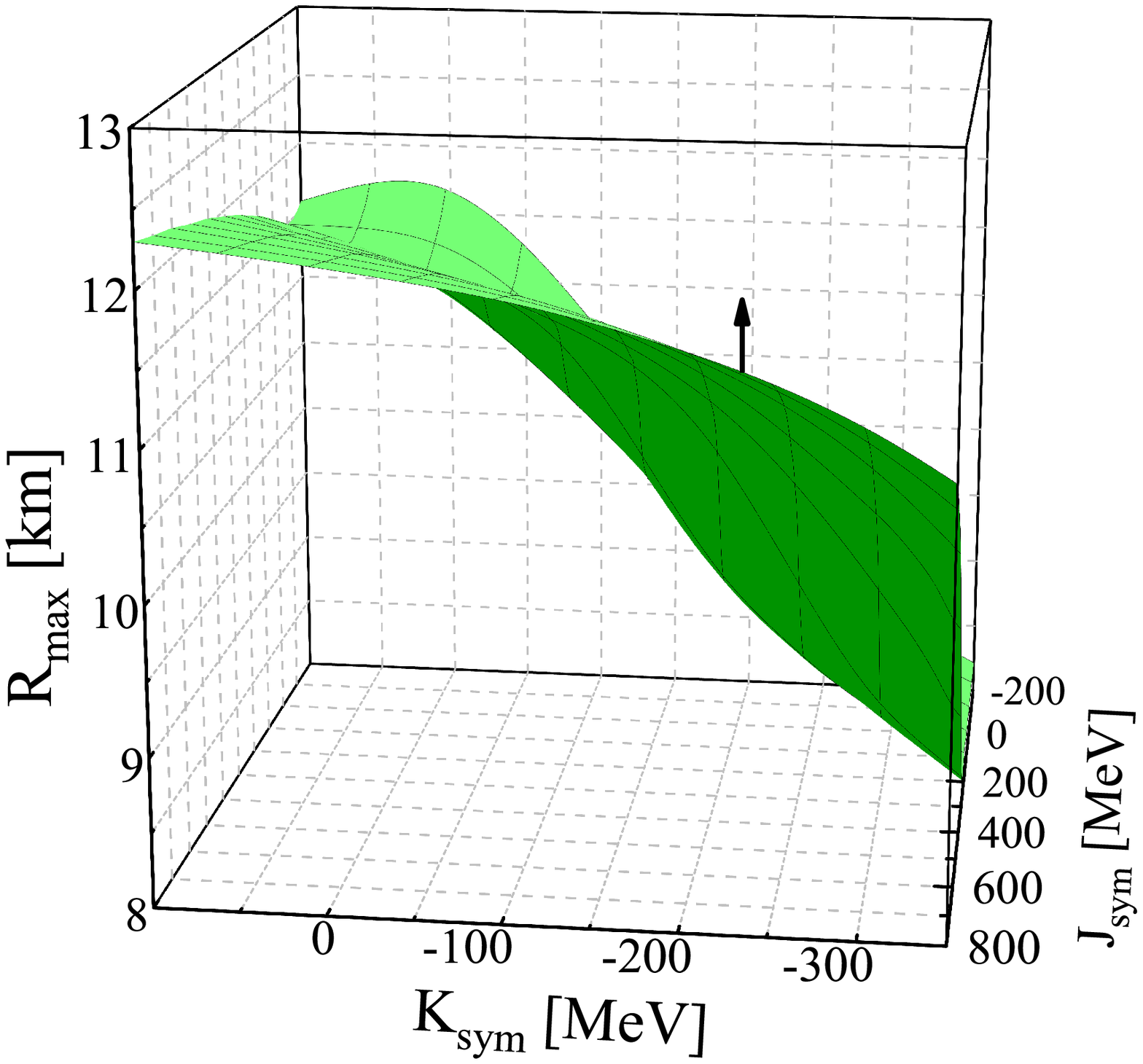}
 }
 \vspace{-0.9cm}
  \caption{The central density (left panel) and radius of a NS with the maximum mass of 2.01 M$_\odot$ as functions of $K_{\rm sym}$ and $J_{\rm sym}$ parameters of the symmetry energy.}\label{RHOc}
\end{figure*}

The $npe\mu$ model outlined above can be considered as the minimum model for NSs. Other particles and new phases of matter may appear especially in massive NSs. However, there are many uncertainties and unknowns associated with these new particles and phases. In particular, to our best knowledge, there is currently no community consensus regarding the existence and type of a hadron-quark phase transition but a hadron-quark duality in understanding properties of NSs. For example, strong evidences for a hadron-quark phase transition were found for massive but not canonical NSs in  \citet{Ann20}. However, a very recent analysis including NICER's latest observation for PSR~J0740+6620 disfavours the presence of a strong first-order phase transition from nuclear matter to exotic forms of matter, such as quark matter, inside NSs \citep{Pang21}. 

The Bayesian analysis of canonical NS data in \citet{Xie-QM} using the above metamodel EOS for hadronic matter and the constant speed of sound (CSS) model for quark matter \citep{CSS} strongly favors a large quark core even in canonical NSs with the most probable hadron-quark transition density at $\rho_t/\rho_0=1.6^{+1.2}_{-0.4}$ at 68\% confidence level. Similarly, low hadron-quark transition densities were also found very recently in two independent Bayesian analyses using similar NS data \citep{Tang21,AngLi21}. While the PDFs of some of the six hadronic EOS parameters in Equations (\ref{E0para}) and (\ref{Esympara}) we extracted with and without considering the first-order hadron-quark phase transition are mildly different, they are all compatible with current constraints within the uncertain ranges of these parameters discussed earlier. In particular, as shown in Figure 12 in \citet{Li21}, while the 68\% confidence boundaries of $E_{\rm{sym}}(\rho)$ at suprasaturation densities become wider when the phase transition is considered as three more parameters in the CSS model are included in the Bayesian analyses of the same set of NS data, existing constrains from nuclear theories, terrestrial experiments, and astrophysical observations can not distinguish the results obtained with and without considering the hadron-quark phase transition. Thus, there is clearly a hadron-quark duality in describing NS observables similar to the situation in relativistic heavy-ion collisions. Namely, purely hadronic models and the ones considering a hadron-quark phase transition can describe equally well all observations within currently acceptable/existing uncertainties of models/observations. Since we are not trying to find ways to break this duality, the $npe\mu$ model is sufficient for the purposes of this work. In our opinion, the choice of using the $npe\mu$ model is reasonable given the current situation of the field although it explores a limited EOS parameter space for NSs.

Unlike in Bayesian statistical analyses normally using the Markov-Chain Monte Carlo sampling, our direct inversions of NS observables in the 3D high-density EOS parameter space are done by brute force. Namely, using the metamodel EOS outlined above, for a given observable of NSs we loop through the 3-dimensional high-density EOS parameter space spanned by the $K_{\rm{sym}}$, $J_{\rm{sym}}$, and $J_0$ parameters within their currently known uncertainties given earlier. At each step of the three loops, we call the Tolman-Oppenheimer-Volkov (TOV) solver.
In looping through the entire high-density EOS parameter space, we also require all accepted EOSs to satisfy the causality condition, retain the dynamical stability through out the entire NS as well as a positive pressure at the crust-core transition density. All accepted EOSs giving the same NS observable will be presented by a constant observable surface in the $K_{\rm{sym}}$-$J_{\rm{sym}}$-$J_0$ EOS parameter space. We emphasize that every point on the constant surface represents a unique EOS and all EOSs on the surface have the same probability to reproduce the specified NS observable. Crosslines of two observable surfaces and/or those between an observable surface and the causality surface will set boundaries for the acceptable high-density EOS parameter space.

As an example, shown in Figure\ \ref{NICER2L60LIGO} are the direct inversions of (1) the upper limit of the tidal deformability of NSs in GW170817 \citep{LIGO18} $\Lambda_{1.4}= 580$ at 90\% confidence level (yellow surface) and  $\Lambda_{1.4}= 427$ at 68\% confidence level (pink surface), respectively, (2) the causality surface (blue surface) on which the speed of sound equals the speed of light ($v^2_s=dP/d\epsilon=c^2$) at the central density of the most massive NS supported by the nuclear pressure at each point with the specific EOS there \citep{Zhang19}, and (3) the constant maximum mass of $M$=2.01 M$_\odot$ which is the lower limit of the newly updated mass of PSR~J0740+6620 \citep{Fonseca21,Miller21,Riley21}. In this work, we use this mass as the {\it minimum} maximum mass of NSs that all EOSs have to be able to support. Namely, all EOSs have to generate stable mass-radius curves having maximum masses higher than or equal to 2.01 M$_\odot$. These constant surfaces were found by calling the TOV solver at each step in looping through the $K_{\rm{sym}}$-$J_{\rm{sym}}$-$J_0$ EOS parameter space in the ranges specified by the three axises of the plot.

The $M=2.01$ M$_\odot$ (green) surface together with the causality surface (blue) limit the allowed range of the $J_0$, thus the high-density SNM EOS. They are essentially vertical to the $J_0$ axis when the $E_{\rm{sym}}(\rho)$ is stiff with large positive $K_{\rm{sym}}$ and $J_{\rm{sym}}$ values. This is because when the symmetry energy is stiff, as shown in Figure \ref{Esymplot}, the isospin asymmetry $\delta=1-2X_p$ at $\beta$-equilibrium is small (Quantitatively from solving the equation $\mu_n-\mu_p=\mu_e=\mu_\mu\approx4\delta E_{\rm{sym}}(\rho)$. Qualitatively, it can be understood simply from minimizing the $E_{\rm{sym}}(\rho)\cdot \delta^2$ term in Equation (\ref{eos})). The nuclear pressure is then dominated by the contribution from SNM EOS. However, when the $E_{\rm{sym}}(\rho)$ becomes super-soft with large negative
$K_{\rm{sym}}$ and/or $J_{\rm{sym}}$ values, nuclear matter becomes very neutron-rich with large $\delta$ values approaching 1. Then, the symmetry energy contribution to nuclear pressure becomes large and can be even negative. Thus to support the same NS mass  $M=2.01$ M$_\odot$, the SNM EOS has to become stiffer with higher $J_0$ values, leading to the bending up of the $M=2.01$ M$_\odot$ surface at the right-back corner where the $K_{\rm{sym}}$ and/or $J_{\rm{sym}}$ are largely negative. To see the maximum density reached in this NS of mass $M$=2.01 M$_\odot$ and its radius, shown in Figure \ref{RHOc} are its central density (left panel) and radius (right panel) as functions of $K_{\rm sym}$ and $J_{\rm sym}$ parameters of the symmetry energy. The central density ranges between about $(5.3\sim 9.5)\rho_0$ while the corresponding radius is between about $(8.5\sim 12.3)$ km. As we shall discuss next, the absolutely maximum mass of non-rotating NSs is determined by the causality surface. If a NS with mass $M=2.01$ M$_\odot$ is below the maximum mass supported by a given EOS, its central density will be lower and its radius larger than the constant surfaces shown in Figure \ref{RHOc} as indicated by the two black arrows.

For most of the $J_{\rm{sym}}$ values the causality surface bends downward as the $K_{\rm{sym}}$ parameter goes from positive to large negative values. As discussed in detail in \citet{Zhang19}, this is because both the NS maximum mass $M_{\rm{cau}}$ on the causality surface that the EOS can support and the corresponding radius $R_{\rm{cau}}$ decrease when the $E_{\rm{sym}}(\rho)$ becomes more soft as the $K_{\rm{sym}}$ decreases. Since the maximum density $\rho_{\rm{max}}$ reached in this maximum mass NS scales with $M_{\rm{cau}}/R^3_{\rm{cau}}$, it increases very quickly with the decreasing curvature $K_{\rm{sym}}$ towards super-soft $E_{\rm{sym}}(\rho)$. Therefore, the causality condition $v_s=c$ can be reached at much smaller $J_0$ values as the $K_{\rm{sym}}$
decreases. We notice that the variation of the causality surface with the skewness parameter $J_{\rm{sym}}$ is more complicated as its effects on both the maximum mass and the corresponding radius are non-monotonic depending on the SNM EOS parameter $J_0$.

Continuing the discussions of the results shown in Figure\ \ref{NICER2L60LIGO}, the crossline between the causality surface and the  $\Lambda_{1.4}= 580$ (427) surface determines the upper limit of $E_{\rm{sym}}(\rho)$ at 90\% (68\%) confidence level. Since the crosslines between the causality surface and the upper radius limits from NICER's observation of PSR J0740+6620 are on the left of the $\Lambda_{1.4}=580$ (90\% confidence) surface, they are not shown here. We emphasize that before considering the lower radius limits from NICER's observation of PSR J0740+6620, the crossline between the causality surface and the NS {\it minimum} maximum mass $M=2.01$ M$_\odot$ sets the lower boundary for the high-density symmetry energy. Both the 68\% upper and lower boundaries of $E_{\rm{sym}}(\rho)$ from these crosslines will be compared to the new 68\% lower boundaries set by the lower radius boundaries from the two independent analyses of NICER's observation of PSR J0740+6620 in the next section. 

\section{The power of radius measurement of massive neutron stars on constraining high-density nuclear EOS}\label{Results}
\begin{figure}
  \centering
   \resizebox{0.48\textwidth}{!}{
  \includegraphics[bb=30 70 550 570]{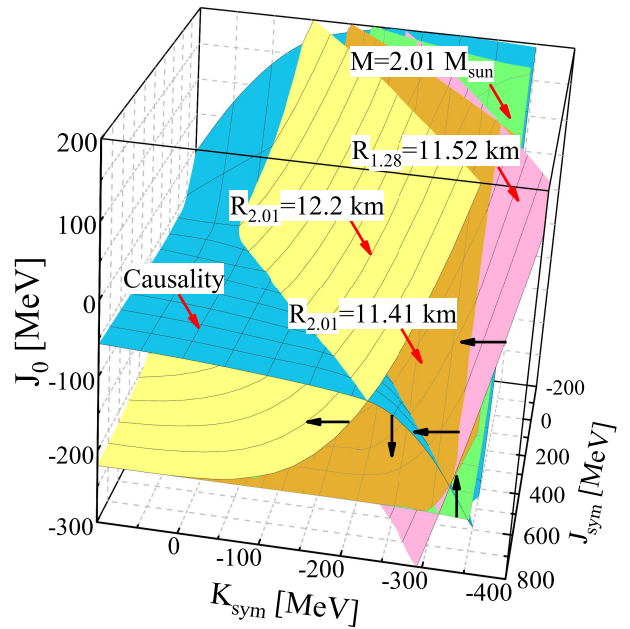}
  }
   \caption{ Similar to Figure \ref{NICER2L60LIGO} but replacing the tidal deformations with NICER's radius lower limit $R_{1.28}=11.52$ km (pink surface) for PSR J0030+0451 \citep{Riley2019} as well as $R_{2.01}=11.41$ (orange surface) km \citep{Riley21} and $R_{2.01}=12.2$ km (yellow surface) \citep{Miller21} for PSR J0740+6620.}\label{NICER2L60}
\end{figure}

Shown in Figure\ \ref{NICER2L60} are the inversions of the NICER's radius lower limit $R_{1.28}=11.52$ km (pink surface) for PSR J0030+0451 \citep{Riley2019} as well as $R_{2.01}=11.41$ (orange surface) km \citep{Riley21} and $R_{2.01}=12.2$ km (yellow surface) \citep{Miller21} for PSR J0740+6620 together with the causality and the {\it minimum} maximum mass $M=2.01$ M$_\odot$ surfaces in the 3D high-density EOS parameter space.
While the $R_{1.28}=11.52$ km surface is on the right of the $R_{2.01}=11.41$ km surface, it is shown here to make a comparison.
Similarly, while the $M$=2.01 M$_\odot$ surface is below the $R_{2.01}=11.41$ km surface and thus does not provide the lower boundary of $E_{\rm sym}(\rho)$ any more, it is shown here to compare the constraint on $E_{\rm sym}(\rho)$ obtained when only the mass of PSR J0740+6620 is measured with that when both its mass and radius are measured. The black arrows indicate the direction to the allowed EOS space.

Several interesting and important observations can be made from comparing the constant surfaces of NS observables and the causality condition as well as their crosslines shown in Figure\ \ref{NICER2L60}. In particular, we emphasize the following points:
\begin{itemize}

\item
The power in limiting the EOS parameter space using simultaneously the knowledge about both the mass and radius of a massive NS can be seen clearly from comparing the $M=2.01$ M$_\odot$ surface and the $R_{2.01}=12.2$ or 11.41 km surface. The separation between the $R_{2.01}=12.2$ km and $R_{2.01}=11.41$ km surfaces indicate the importance of further reducing the uncertainty of the radius measurement. For the discussions here we use the
$R_{2.01}=12.2$ km surface, while the lower boundaries of $E_{\rm{sym}}(\rho)$ using both radii will be obtained and compared later.
There is a huge gap in the direction of $J_0$ between the two surfaces in the area where the $E_{\rm{sym}}(\rho)$ is super-soft. The gap closes gradually as the $E_{\rm{sym}}(\rho)$ becomes super-stiff towards the front-left corner. This is understood again because the pressure is dominated by the SNM EOS when the $E_{\rm{sym}}(\rho)$ is super-stiff and the corresponding isospin asymmetry $\delta$ vanishes. In this region of the high-density EOS parameter space, the only mechanism to change both the NS mass and its radius simultaneously is through the variation of $J_0$.

On the other hand, it is seen that in the space where the $E_{\rm{sym}}(\rho)$ is super-soft, the $R_{2.01}=12.2$ km surface is almost vertical while the $M=2.01$ M$_\odot$ surface is still rather flat (which can be better seen in Figure \ref{NICER2L60LIGO}), meaning that the variation of $J_0$ has essentially no effect on the radius of PSR J0740+6620 in this EOS parameter region. Most importantly, the crossline between the  $R_{2.01}=12.2$ km surface and the causality surface sets the new lower boundary of $E_{\rm{sym}}(\rho)$. Compared to the lower boundary set by the crossline between the $M=2.01$ M$_\odot$ and causality surfaces shown in Figure \ref{NICER2L60LIGO}, it moved upward (become stiffer) significantly as we shall discuss more quantitatively below. Therefore, the most important effect of knowing simultaneously both the radius and mass of the most massive NS observed is in pinning down the high-density symmetry energy especially its lower boundary.

\item
The power in limiting the EOS parameter space of knowing the radius of a massive NS compared to that of a light one can be clearly seen by comparing the $R_{2.01}=12.2$ km surface and the $R_{1.28}=11.52$ km surface. The latter is rather vertical because in the whole EOS space considered, all values of $J_0$
can support a NS of mass $M=1.28$ M$_{\odot}$ that has a relatively low central density and the $J_0$ has almost no effect on the radii of light NSs, in contrast to the situation for the most massive NS observed so far as we discussed above. The separation between these two surfaces clearly indicates the strong sensitivity of NS radius to the variation of $E_{\rm{sym}}(\rho)$. Since both surfaces are the lower radius limits of the two NSs considered, it is obvious that the $R_{2.01}=12.2$ km surface provides a more tight lower limit for the $E_{\rm{sym}}(\rho)$, while the crossline of the $R_{1.28}=11.52$ km surface with either the causality or the $M=2.01~M_{\odot}$ surface gives a lower boundary very close to that from the crossline between the causality and $M=2.01$ $M_{\odot}$ surfaces.

\end{itemize}

\begin{figure}
  \centering
   \resizebox{0.48\textwidth}{!}{
  \includegraphics{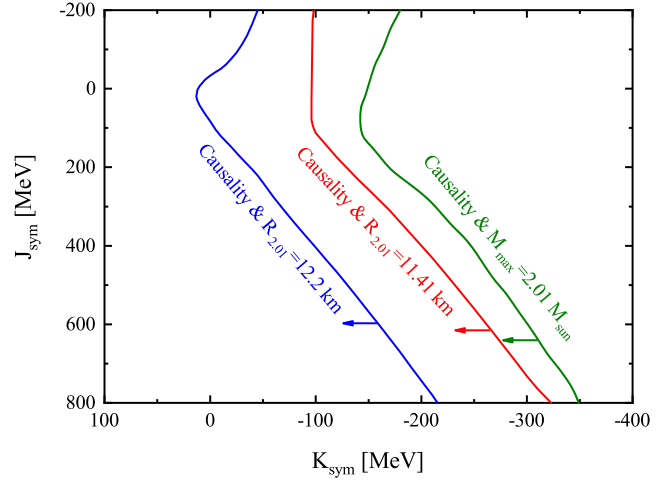}
  }
  \caption{The projection of the crossline between surfaces of causality condition and $M_{\rm max}=2.01$ M$_\odot$ (green line), $R_{2.01}=11.41$ km (red line) and $R_{2.01}=12.2$ km (blue line), respectively,  on the $K_{\rm sym}-J_{\rm sym}$ plane for $L=58.7$ MeV. The arrows point to the directions satisfying the corresponding constraints. }\label{KsymJsymL60}
\end{figure}

\begin{figure}
  \centering
   \resizebox{0.48\textwidth}{!}{
  \includegraphics{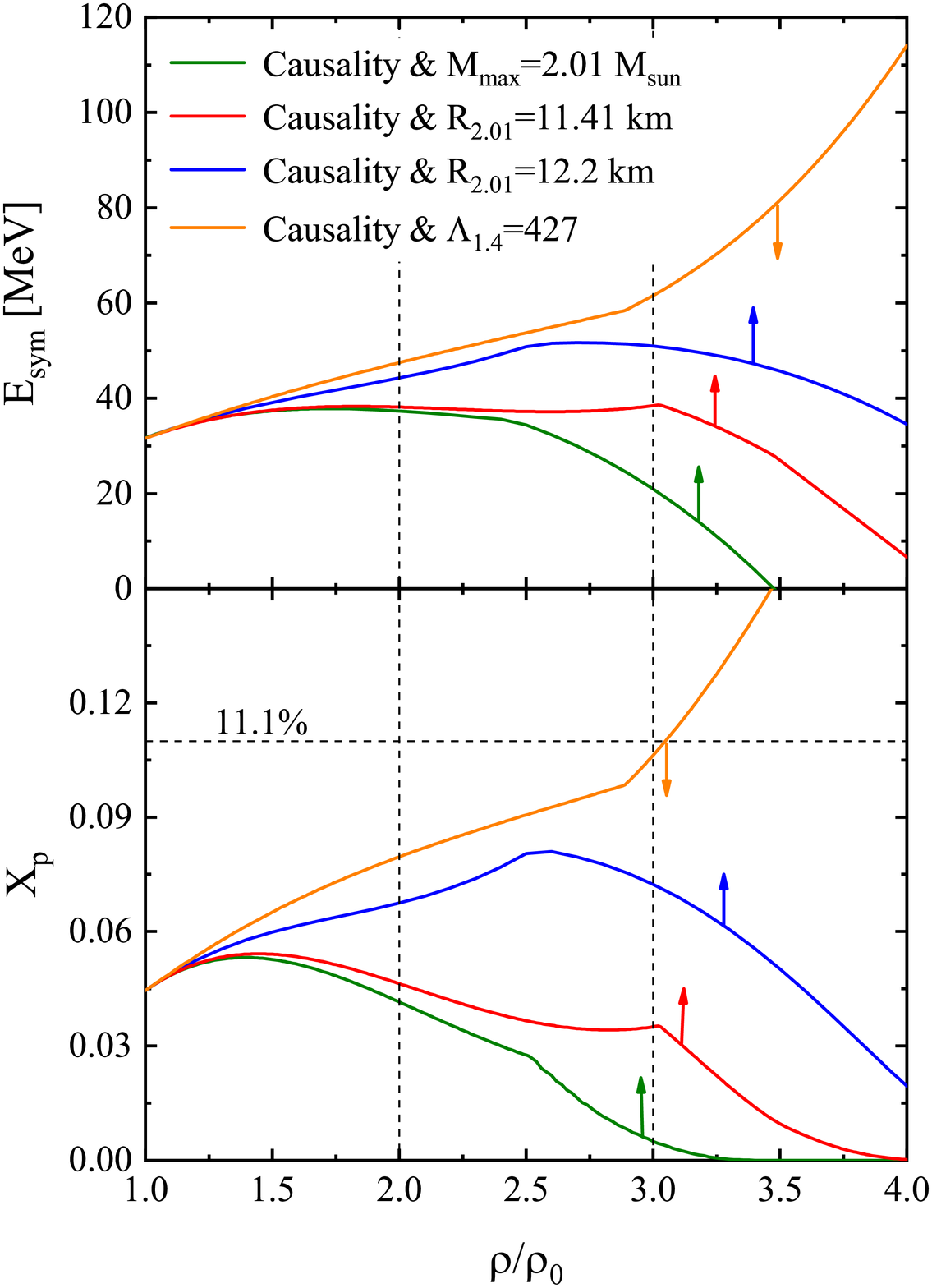}
  }
  \caption{Upper panel: The constrained lower limit of symmetry energy as a function of density for $L=58.7$ MeV when only mass is observed (green line) or both mass and radius are observed (red and blue lines). The constrained upper limit from $\Lambda_{1.4}=427$ is shown as orange line. Lower panel: The corresponding lower limit of proton fraction as a function of density. The horizontal line shows the critical proton fraction 11.1\% for the direct URCA process to occur. The arrows point to the directions satisfying the corresponding constraints.}\label{EsymXp}
\end{figure}

We now turn to the 68\% confidence boundaries of the high-density symmetry energy from the crosslines discussed above. The function $E_{\rm{sym}}(\rho)$ is determined by the $K_{\rm sym}$ and $J_{\rm sym}$ parameters when the $L$ and $E_{\rm sym}(\rho_0)$ are fixed at their most probable values given earlier.  Shown in Figure \ref{KsymJsymL60} are projections of the crossline between surfaces of causality condition and $M_{\rm max}=2.01$ M$_\odot$ (green line), $R_{2.01}=11.41$ km (red line), and $R_{2.01}=12.2$ km (blue line), respectively,  on the $K_{\rm sym}-J_{\rm sym}$ plane for $L=58.7$ MeV. As discussed above, these are the lower boundaries.
The arrows point to the directions satisfying the corresponding constraints. We note that along these boundaries the values of $J_0$ are different as shown in Figures \ref{NICER2L60LIGO} and \ref{NICER2L60}. To see again effects of simultaneously measuring the mass and radius of massive NSs, it is interesting to note that the boundary from the crossline between the causality and $M=2.01~M_{\odot}$ surfaces sets the lower limit of $K_{\rm sym}$ at -349 MeV. The information about the radius lower limit moves the latter significantly. For example, the lower radius limit $R_{2.01}=12.2$ km moves it to about -215 MeV, thus reducing the
uncertainty range of $K_{\rm sym}$ by a factor of about 1.6. However, it is seen that the difference due to the systematic error in analyzing NICER's radius measurements is quite large.
Moreover, the $J_{\rm sym}$ parameter is still not constrained by the NS observations available. Hopefully it can be constrained by future radius measurements of even more massive NSs, messengers directly from the core of NSs, e.g., neutrinos, postmerger high-frequency gravitational waves and/or properties of hyper/super-massive remnants following mergers of two NSs.

Using the Equation (\ref{Esympara}), the 68\% confidence boundaries in the $K_{\rm sym}-J_{\rm sym}$ plane can be translated into constraints for the symmetry energy $E_{\rm sym}(\rho)$ as shown in the upper panel of Figure \ref{EsymXp}. For a comparison, the upper 68\% confidence boundary from the crossline between the $\Lambda_{1.4}=427$ and causality surfaces is shown as the orange line. In translating the boundaries in the  $K_{\rm sym}-J_{\rm sym}$ parameter plane to the lower/upper boundaries of $E_{\rm{sym}}(\rho)$, since each point on the boundary generates a new function $E_{\rm{sym}}(\rho)$ we have to compare all functions generated along a given boundary to find the lowest/highest $E_{\rm{sym}}(\rho)$ value at each density $\rho$. Then, linking all lowest/highest values at all densities gives us the lower/upper boundary of symmetry energy as a function of density. Because of the finite bin size in density used in the translation, the resulting $E_{\rm{sym}}(\rho)$ functions are not very smooth at all densities.
Clearly, the measured radius of the most massive NS observed so far improves significantly our knowledge about nuclear symmetry energy especially at densities higher than about $2.5\rho_0$. More quantitatively, the lowest symmetry energies at $2\rho_0$ and $3\rho_0$ and the corresponding proton fractions from the three crosslines are listed in Table \ref{tab:1}.  Again, the lower radius limits from analyzing NICER's radius measurement of PSR J0740+6620 significantly increases the allowed lowest value of symmetry energy at these two densities. 

As mentioned earlier, the density profile of proton fraction $X_P(\rho)$ determined uniquely by the $E_{\rm sym}(\rho)$ is a critical quantity for addressing several interesting NS physics issues. The constrained proton fraction $X_P(\rho)$ is shown in the lower panel of Figure \ref{EsymXp}. The direct URCA process is predicted to occur once the proton fraction exceeds the critical fraction of about $11.1\%$ (shown as the horizontal dashed line) \citep{Kla06}. The results of our analyses indicate that the direct URCA process is disfavoured below about $3\rho_0$ but remains uncertain at higher densities.
\begin{table}[h]
	\centering
	\caption{The 68\% confidence lower limits of symmetry energy and proton fraction at 2$\rho_0$ and 3$\rho_0$ from the crosslines of the causality condition and the three NS observables indicated in the first column.}
	\begin{tabular}{c|cc|cc}\hline\hline
	    Causality \& & $E_{\rm sym}(2\rho_0)$ & $E_{\rm sym}(3\rho_0)$ & $X_p(2\rho_0)$& $X_p(3\rho_0)$ \\
        & (MeV) & (MeV) & (\%) & (\%)\\
    \hline
    $M_{\rm max}=2.01$ M$_\odot$ & 37.32 & 20.96 & 4.1 & 0.5\\
    $R_{2.01}=11.41$ km & 38.16 & 38.46 & 4.6 & 3.5 \\
    $R_{2.01}=12.2$ km & 44.26 & 50.96 & 6.7 & 7.2 \\\hline
	\end{tabular}
	\label{tab:1}
\end{table}

Moreover, the possible formation of proton polarons and several related phenomena as well as the trigger of isospin separation instability in the core of NSs all depend critically on the proton fraction at high densities\citep{Kut93,Kut94,Kubis1,Szm06}. One common condition is that the symmetry energy has to start decreasing with increasing density above certain critical density, leading to the gradual disappearance of protons in the core of NSs. From Figure \ref{EsymXp}, it is seen that the lowest boundary (the green line determined by the crossline between the causality and the $M=2.01~M_{\odot}$ surfaces) of $X_p$ permits this at densities higher than about $2\rho_0$ when only the mass $M=2.01~ M_{\odot}$ is observed for PSR J0740+6620.
In fact, many phenomenological and microscopic models predicted high-density behaviors of the $E_{\rm sym}$ and the corresponding $X_p$ similar to the dashed line, see, e.g., the 10 realistic EOSs used in \citet{Szm06} in studying the proton polaron formations in NSs. Interestingly, however, the lower boundary of $X_p$ from using the lower radius limit $R_{2.01}=12.2$ km or 11.41 km disfavours  the formation of proton polarons at least upto about $3.0\rho_0$. While it is still more likely at higher densities, our present model assuming NSs are made of $npe\mu$ matter may break down and thus can not be used reliably to make predictions at higher densities.

\begin{figure}
  \centering
    \resizebox{0.45\textwidth}{!}{
  \includegraphics[bb=30 60 550 570]{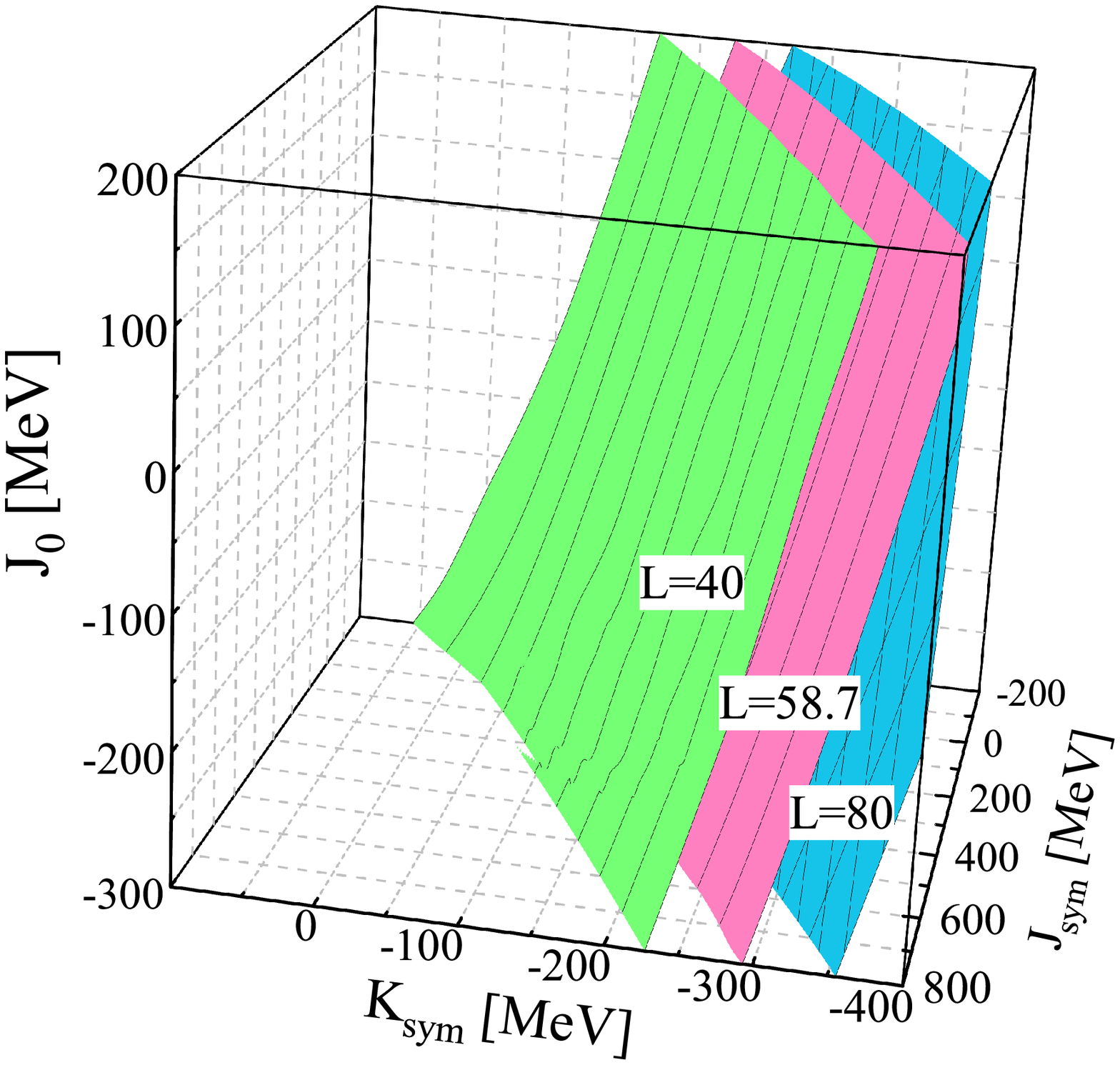}
  }
   \resizebox{0.45\textwidth}{!}{
  \includegraphics[bb=30 60 550 570]{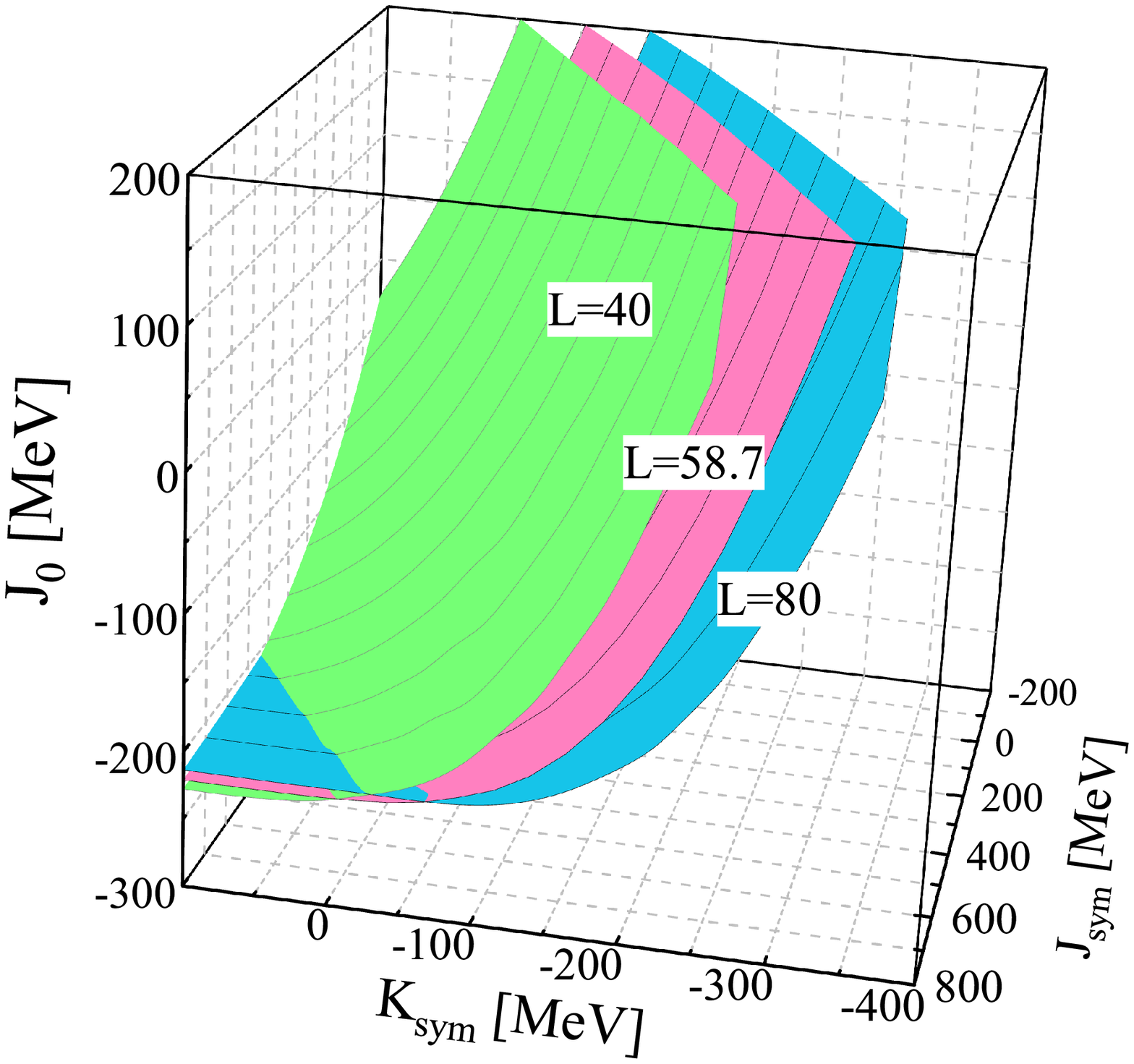}
  }
  \caption{Comparisons of the constant surfaces of $R_{1.28}=11.52$ km (upper panel) and $R_{2.01}=12.2$ km (lower panel) for $L=40$, $58.7$, and $80$ MeV, respectively.}\label{LLcompare}
\end{figure}
\section{Effects of the remaining uncertainty of the slope parameter $L$ of nuclear symmetry energy}
\label{Sec4}
\begin{figure}
  \centering
   \resizebox{0.48\textwidth}{!}{
  \includegraphics[bb=0 0 550 570]{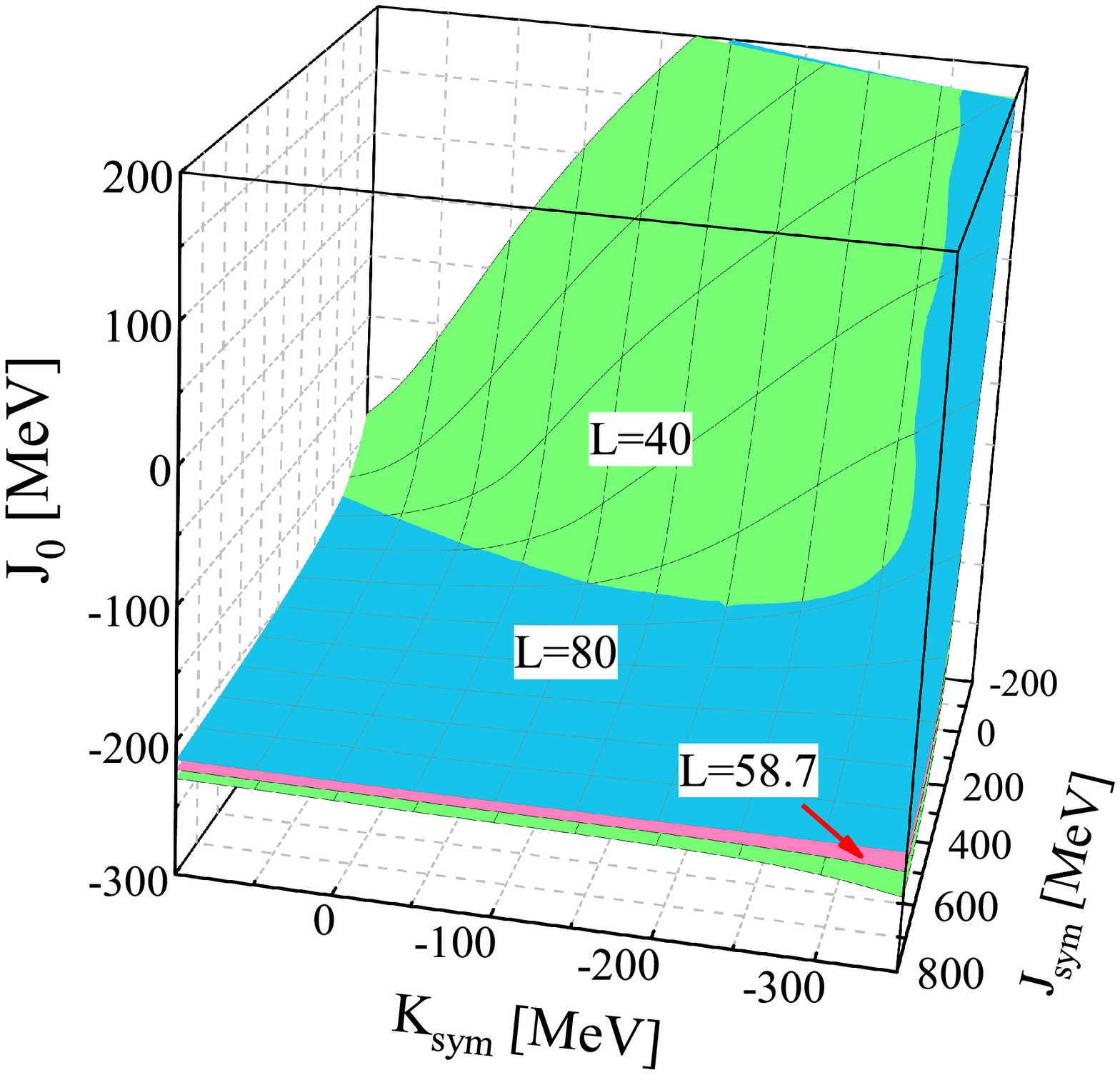}
  }
     \resizebox{0.48\textwidth}{!}{
  \includegraphics[bb=0 0 550 570]{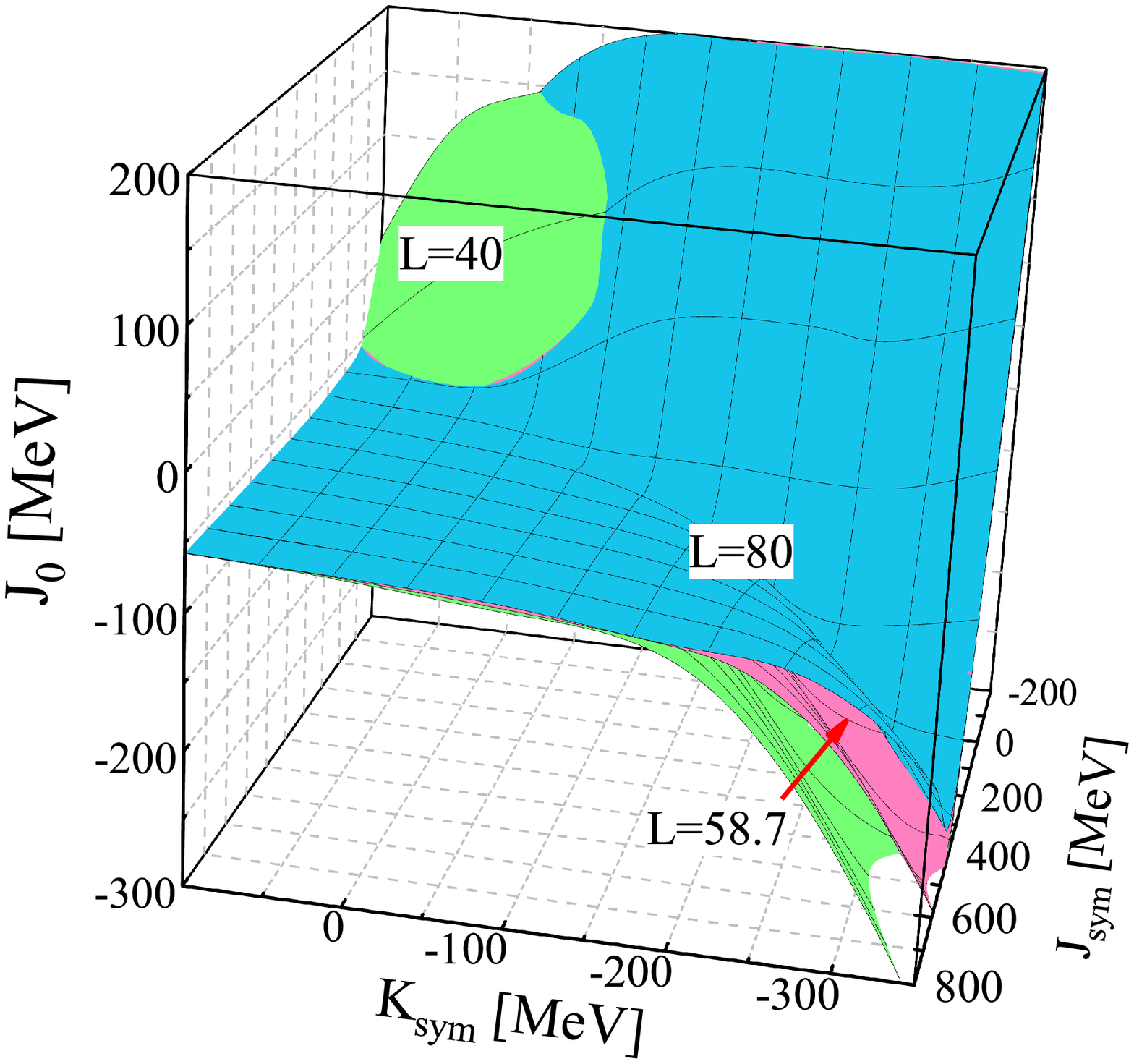}
  }
  \vspace{-1cm}
  \caption{Comparisons of the constant surfaces of mass $M=2.01$ M$_\odot$ (upper panel) and causality condition (lower panel) for $L=40$, $58.7$, and $80$ MeV, respectively.}\label{LM214compare}
\end{figure}

\begin{figure}
  \centering
   \resizebox{0.45\textwidth}{!}{
  \includegraphics{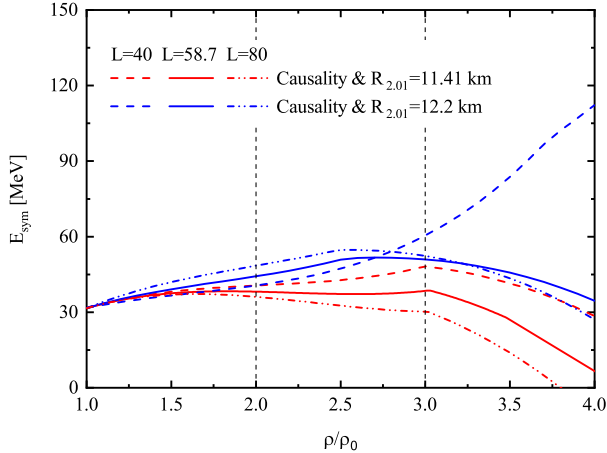}
  }
  \caption{The combined constraints on the lower limit of symmetry energy for $L$ = 40 (dash lines), 58.7 (solid lines), and 80 (dash-dot-dot lines) MeV. }\label{EsymbandCom}
\end{figure}

In the above discussions, the slope $L$ of symmetry energy is set at its fiducial value of $L=58.7$ MeV, while it is known that it still has an uncertainty of about 20 MeV based on the latest survey of available analyses in the literature \citep{Li21}.
The $L$ parameter characterizes mainly the symmetry energy around $\rho_0$. It is well known that $L$ has strong imprints on the
radii and tidal deformations of canonical NSs. However, it is unclear how it may affect the radii and masses of massive NSs as well as the causality surface. Most importantly, how does $L$ affect the lower boundaries of the symmetry energy we extracted above? To address these questions we have re-calculated everything with $L=40$ and 80 MeV and compared the results with the ones we obtained above with $L=58.7$ MeV.
As examples, shown in Figure \ref{LLcompare} are comparisons of the constant surfaces of $R_{1.28}=11.52$ km (upper panel), and $R_{2.01}=12.2$ km (lower panel) with $L=40$, $58.7$, and $80$ MeV, respectively. As expected, the increase of $L$ shifts the surface $R_{1.28}=11.52$ km to smaller $K_{\rm sym}$ values. For the light NSs, when $L$ becomes larger, $K_{\rm sym}$ has to become smaller to reproduce the same radii. This is consistent with our previous findings shown in \citet{Zhang19a}. Interestingly, however, the above expectation is also true for the radii of massive NSs only in the space where the symmetry energy is super-soft as indicated by the surfaces of $R_{2.01}=12.2$ km for different $L$ in the lower panel. In the region where the symmetry energy is super-stiff, the $L$ is seen to have little effect on the $R_{2.01}=12.2$ km surface. As we discussed earlier, when the symmetry energy is stiff (larger $K_{\rm sym}$ and/or $J_{\rm sym}$), the pressure is dominated by the SNM EOS, namely the $J_0$ parameter. Then both the mass and radius are mainly determined by the $J_0$ with little influence from the symmetry energy. Moreover, since the $K_{\rm sym}$ and $J_{\rm sym}$ are both so big in the super-stiff region, the $L$ parameter thus has little effect on the $R_{2.01}=12.2$ km surface when the symmetry energy is super-stiff. For similar reasons, as shown in Figure \ref{LM214compare}, the variation of $L$ has little effect on the constant surface of $M=2.01$ M$_\odot$ and the causality surface, except in the space where the symmetry energy is super-soft.

Since the crosslines determining the boundaries of $E_{\rm sym}$ are not in the super-soft region, effects of $L$ on these boundaries are expected to be small or moderate. Summarized in Figure \ref{EsymbandCom} are constraints on the 68\% confidence lower boundaries of symmetry energy from the two analyses of the NICER data using $L=40$, $58.7$, and $80$ MeV, respectively. It is seen that below about 3$\rho_0$ the variation of the lower boundary of $E_{\rm sym}$ due to variation of $L$ by $\pm 20$ MeV is compatible with the variation due to the systematic error of NICER's radius analysis.

\section{Summary and conclusions}
In summary, by directly inverting several NS observables in the high-density EOS parameter space we have shown that the lower radius limits from NICER's very recent observation of PSR J0740+6620 set a much tighter lower boundary than previously known for nuclear symmetry energy in the density range of $\sim (1.0-3.0)\rho_0$. The super-soft symmetry energy leading to the formation of proton polarons and the related phenomena predicted to occur in this density region in the core of NSs is clearly disfavoured by NICER's radius measurement for the most massive NS observed so far. 

Our work has two major caveats. Firstly, we did not consider any kind of phase transition although the central density may reach far above the normally expected hadron-quark transition density in PSR J0740+6620. 
Naturally, the additional flexibility from adding a phase transition widens the range of hadronic matter EOS parameters that agrees with the lower radius bound of PSR J0740+6620. Moreover, the physical meaning of nuclear symmetry energy ends at the hadron-quark phase transition. Secondly, our EOS parameterizations ends at the 
$\chi^3$ terms. Including higher-order terms adds more flexibilities of describing the high-density behavior of nuclear EOS. Since our direct inversions of NS observables by brute force are limited to 3D, effects of high-order terms are not investigated in this work. These may introduce some systematic uncertainties in our results. Our findings should thus be understood with these cautions in mind. Nevertheless, the relative positions of the lower boundaries of high-density symmetry energy from using different NS observables extracted from the same approach are physically sound and qualitatively reliable.

\section*{Acknowledgement} We would like to thank Wen-Jie Xie for helpful discussions. This work is supported in part by the U.S. Department of Energy, Office of Science, under Award Number DE-SC0013702, the CUSTIPEN (China-U.S. Theory Institute for Physics with Exotic Nuclei) under the US Department of Energy Grant No. DE-SC0009971, the National Natural Science Foundation of China under Grant No. 12005118, and the Shandong Provincial Natural Science Foundation under Grants No. ZR2020QA085.

\end{document}